\documentclass[twocolumn,amssymb, nobibnotes, aps, prb,groupedaddress]{revtex4-1}
\usepackage{graphicx}

\begin{document}
\begin{abstract}
Density functional theory (DFT) and many body perturbation theory at the G$_0$W$_0$ level are employed to study the electronic properties of polythiophene (PT) adsorbed on graphene surface. Analysis of charge density difference  shows the substrate-adsorbate interaction leading to a strong physisorption and interfacial electric dipole moment formation. The electrostatic potential displays  a -0.19 eV shift in the graphene work function from its initial value of 4.53 eV, as the result of the interaction. The LDA band gap of the polymer does not show any change, however the energy level lineshapes are modified by the orbital hybridization. The  interfacial polarization effects on the band gap and levels alignment are investigated within G$_0$W$_0$ level and shows notable reduction of PT band gap compared to that of the isolated chain. 
\end{abstract}

\title{Improving the Performance of Polythiophene-based Electronic devices by Controlling the Band Gap in the Presence of Graphene}%

\author{F. Marsusi}%
\email{marsusi@aut.ac.ir}
\affiliation{$^1$Department of Physics, Amirkabir University of Technology, P.O. Box 15875-4413, Tehran, Iran
}
\author{I. A. Fedorov}
\email{marsusi@aut.ac.ir}
\affiliation{$^2$Department of Theoretical Physics, Institute of Fundamental Sciences, Kemerovo State University, 650043, Kemerovo, Russia
}
\email{ifedorov@kemsu.ru}
\author{S. Gerivani}
\affiliation{$^3$Department of Physics, Amirkabir University of Technology, P.O. Box 15875-4413, Tehran, Iran
}
\date{August 10, 2017}%
\maketitle

\section{INTRODUCTION}
Polythiophene (PT) and its derivatives are the most important stable and ease of preparation $\pi$--conjugated semiconductors with a broad spectrum of applications in organic electronics. The applications range from the organic light-emitting diodes (OLED) and displays to solar cells and sensors \cite{Li, Reynolds, Perzon, Ho, Mwaura, Zou}. However, PT--derivatives based electronic devices suffer from relatively large electronic band gap and low carrier mobility which reduce the corresponding quantum efficiency \cite{Kaloni, Zhou}. \\
On the other hand, due to its ballistic charge transport and high electron mobility, graphene expected to be an outstanding new material for the electronic applications. These exceptional features suggest that graphene can be utilized to improve the electronic characteristic and charge transport of polymer--based organic semiconductors. Single and few multilayers graphene are transparent. This property is also of high importance for those devices that the light should enter their active layer. Recently, organic PT-derivatives, poly (3-hexylthiophene (P3HT) and poly(3-octylthiophene) (P3OT) photovoltaic as electron donors and an organic functionalized graphene as electron-accepter has been fabricated \cite{Liu}. It was shown that this composite works well due to the interaction between graphene and polymers \cite{Liu}. Recently, the crystallization of highly regular P3HT on single layer graphene is also reported \cite{Skrypnychuk}. The P3HT polymer film deposited on graphene reported to have a very different distribution of crystallite orientations. Also much higher degree of $\pi-\pi$ stacking perpendicular to the plane of the graphene film, compared to deposition on silicon surface was exhibited \cite{Skrypnychuk}. Graphene is reported as an ideal template for growing ultra-flat organic films with a face-on orientation \cite{Wang}. These preference features are accounted to obtain higher charge transports and efficiencies from PT derivatives \cite{Skrypnychuk}. \\
Therefore, this paper presents a theoretical attempt to unveil and understand theoretically the influence of graphene at the electronic properties of PT. To this end, first the electronic structure of isolated graphene and PT are investigated. Next, the adsorption mechanism and the modification of the electronic structure when PT adsorbed on the graphene surface are inspected.  In order to obtain a comprehensive view, we use three different approaches: (i) local density approximation (LDA), (ii) the Hubbard corrected LDA+U functional and (iii) the many-body perturbation theory at the $G_0W_0$ level. The outcomes are compared with the previous available theoretical and experimental data. As our main purpose, PT is brought close to the graphene surface in face-on orientation. Then it is inspected that how the electronic states of isolated PT are perturbed by the graphene surface using the three mentioned approaches. Van der Waals (vdW) interaction are considered when the binding energy are calculated by semi-local PBE-GGA functional. Our results show that the most stable geometry occurs in the adsorption distance of about 3.4-3.5 \AA. Since a typical adsorbate-substrate spacing in a physisorption is more than 3 \AA, the obtained value is an evidence of a physisorption. Charge density analysis shows no charge is transferred between PT and graphene. However, results show a clear charge density distortion close to the graphene surface, which is a sign of electric dipole formation. Previous many body ab initio studies have clarified that the polarization, even for a physisorbed and weakly coupled interaction, can considerably influence the size of the adsorbate gap \cite{Fu, Puschnig}. It is also well known that local (semi-local) nature of density functional theory (DFT) functional cannot contribute for the polarization effect in the interface, due to self-interaction errors. By ignoring the polarization effects, the LDA PT gap is independent of the underlying substrate, and the only connection with the graphene is done through a weak interaction with slightly change in the electronic states through orbital hybridizations. It is well known that the self-interaction error which appears in the occupied states in the standard DFT with local (or semi-local) exchange-correlation functionals over delocalizes the highest occupied state and pushes it up, therefore reduces the band gap \cite{Puschnig, Lanzillo, Garcia, Neaton}. Subsequently, when a molecule is brought close to the substrate, local (semi-local) functionals exaggerate the density distribution for the added electron or hole. The consequent of the delocalization error is incorrect convex behavior (instead of piecewise constant) of the energy with respect to the number of electrons . As a result, the derivatives of the energy with respect to fractional charge produce incorrect ionization energy and electron affinity and smaller band gap \cite{Mori}. \\
 DFT+U inspired by the Hubbard model as a typical approach may be used to correct the DFT unphysical curvature of the total energy. Our outcomes show that as for many typical conjugated polymers, DFT+U does not improve the band gaps of isolated PT and the adsorbed PT on graphene significantly \cite{Baeriswyl}. In order to describe the dynamical polarization effect of the surrounding electrons, we employed many-body perturbation theory implemented through G$_0$W$_0$ approximation on the top of the DFT calculations. The self-energy $\Sigma$ is given as the product of the non-interacting single-particle Green function $G_0$, and the dynamically screened Coulomb interaction $W_0$, calculated within the plasmon pole approximation. It is well-known that including dynamical electronic correlation in G$_0$W$_0$ approximation mimics the long-range image potential effects of electrons near the interface \cite{Inkson}. Using G$_0$W$_0$, we obtained a relatively strong renormalization of the PT band gap, once it physisorbed on the graphene surface. 
 \section{COMPUTATIONAL DETAILS}\label{sec: computational}
\begin{figure}
\caption{\label{fig1} {(Color online) (a) Top view showing the monoclinic  supercell model describing PT polymer adsorbed on top of the graphene monolayer. For clarity the cell is duplicated in both in-plane directions. (b) Top view showing the unit cell vectors of the hexagonal graphene primitive cell $\textbf{a}_1$ and $\textbf{a}_2$ (blue arrows), as well as the monoclinic supercell basis vectors $\textbf{a}'_1$ and $\textbf{a}'_2$ (red arrows). (c) Illustration of the supercell Brillouin zone along with the high-symmetry points that are selected to define \textbf{k}-vector paths.}}\
\centering
\includegraphics[scale=0.65]{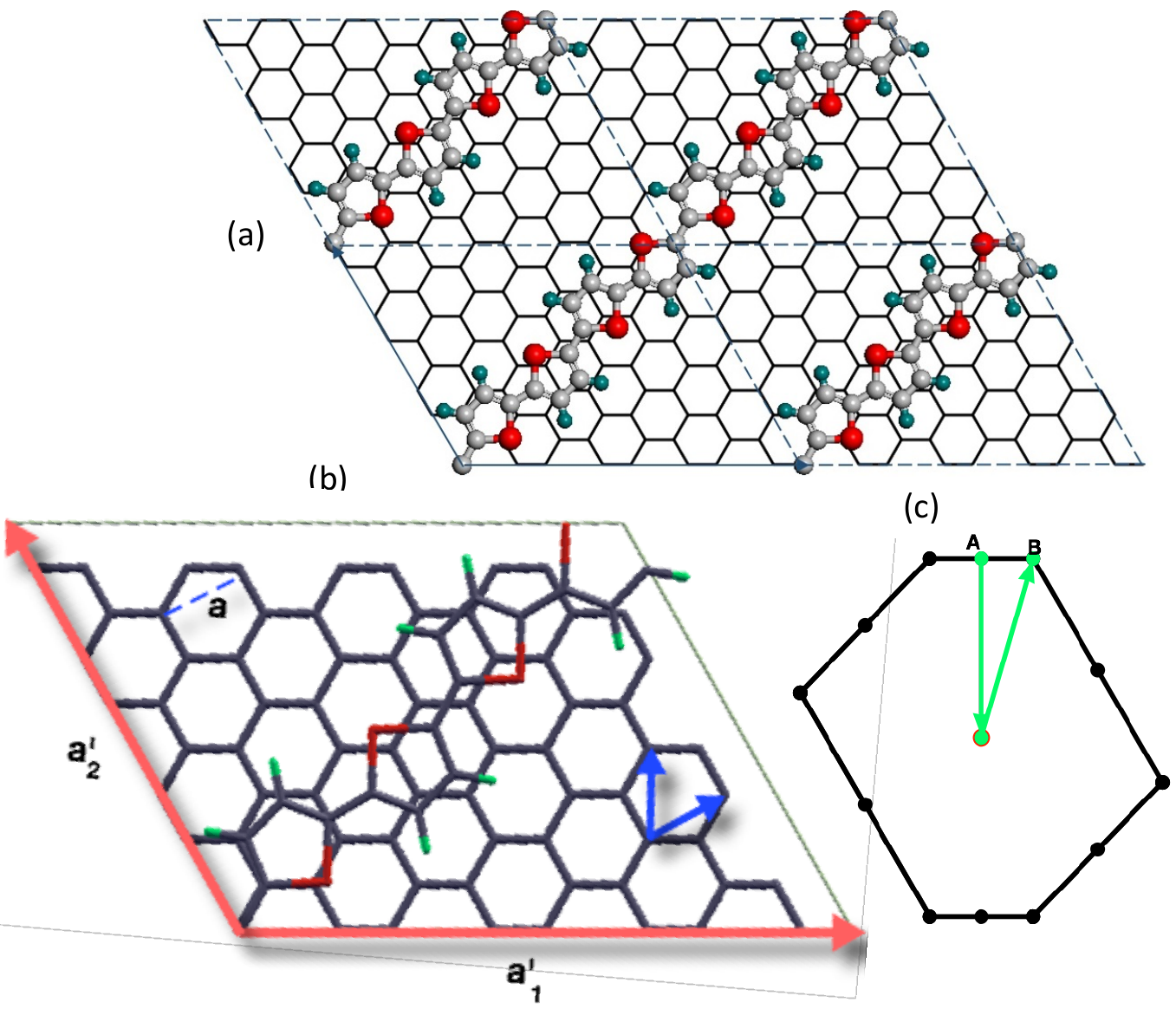} 
\end{figure}

 The formation of a regular PT and graphene layer will be considered, when the lateral PT-PT interactions to be less attractive than the corresponding interaction between PT and the graphene surface. To this end, we should perform the calculations in a supercell with the size large enough to prevent the interactions between PT and its images. In addition, the thiophene oligomer in the supercell must arranged in a way that forming the polymer chain, when the cell subjected to the periodic boundary condition (PBC). At the same time, from the repetition of this unit cell in both two coordinates (x, y), the graphene sheet must be constructed. However, the main axes of the graphene lattice do not commensurate at all with the periodicity of PT. Therefore, constructing PT in the face-on orientation is not a trivial commission. In practice, there are many possible on-face arrangement for the combined system, but regarding the computational limitation was explained, we found that the concurrent assembly can be best done when the oligomer is placed along the base-diameter of the monoclinic supercell shown in Fig. \ref{fig1}(a).

 This arrangement provides us an opportunity to study and analyze the electronic behaviour of PT adsorbed on graphene surface, at the cost of increasing the number of carbon atoms in the graphene cell up to 72 atoms. If we denote the graphene primitive vectors by 
$\textbf{a}_1$ and $\textbf{a}_2$ ($|\textbf{a}_1|$=$|\textbf{a}_2|$=a (1.73 \AA)),
 the supercell vectors assembled in this work can be described by $\textbf{a$^\prime$}_1$
 ($|\textbf{a$^\prime$}_1|=\frac{12a}{\sqrt{3}}=17.15 $~\AA) 
 and $\textbf{a$^\prime$}_2$ ($|\textbf{a$^\prime$}_2|$=$\frac{9a}{\sqrt{3}}$=12.86 \AA). A vacuum region of 14~\AA~along the perpendicular direction is imposed to guarantee a vanishing interaction between periodically repeated images. By this value the total energy versus z--component of the supercell is converged to less than 2 meV.\\
 By repetition of the supercell along the two in-plane vectors, both graphene sheet and PT are built, as shown in Fig. \ref{fig1}(b), while the supercell is large enough to prevent the interaction between PT-PT themselves, which are at the distance of 12.8~\AA~  from each other. The first Brillouin zone (BZ) corresponding to the supercell is shown in Fig. \ref{fig1}(c). The two paths through the three symmetric points of the BZ are shown in Fig. \ref{fig1}(c). These points are selected to investigate the electronic behaviours of the isolated components and combined system. Optimization procedure has been performed within LDA frame work by using plane-wave pseudopotential as implemented in the ABINIT code \cite{ABINIT}. To converge the total energy to within 1 meV a plane-wave cutoff energy of 30 a.u. is required. The Brillouin zone (BZ) are sampled by $4\times4\times1$ Monkhorst-Pack \textbf{k}-vectors \cite{Monkhorst}. LDA Troullier–-Martins (TM) pseudopotentials are used in our calculations \cite{Troullier}. Both volume of the supercell and the position of PT atoms inside it are relaxed within the forces less than 5 meV/\AA. The structure of isolated graphene sheet are relaxed within the same criteria. \\
 The U correction are adopted to the p--orbitals of all atoms in the cell within the projector augmented-wave (PAW) method. The PAW kinetic energy cut-off is converged by 20 a.u. To determine the U and J parameters, we follow a semi-empirical strategy by looking for the optimum values when the band gap is increased to its experimental value.\\

 \begin{figure}
\caption{\label{fig2} {(Color online) (a) The optimized geometry model of PT oligomer used in the supercell explained in the text. Atoms in a hexagonal ring are labelled according to parameters shown Table I.}}
\centering
\includegraphics[scale=0.5]{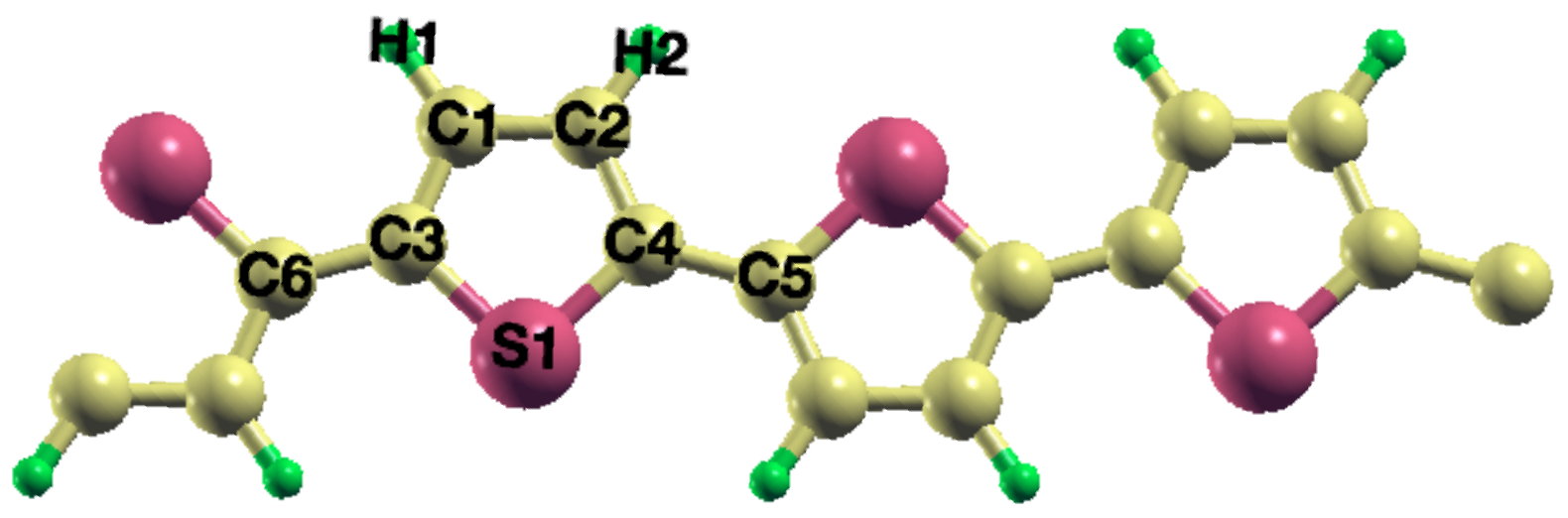} 
\end{figure}

\begin{table}[h!]
\centering
\caption{\label{tab1} Calculated optimized bond lenghts and angles of polythiophene according to atoms labeled in Fig. \ref{fig2}. The  previous available LDA predictions are listed for comparison. All lengths are in \AA~ and angles in degree.}
\begin{tabular}{@{}lll}
\toprule
parameter&LDA&$^a$LDA\\
\hline
C6--C3& 1.42 &1.40\\
C3=C1&1.38& 1.38\\
C1--C2&1.40&\\
C1--H1&1.09&\\
C4--C5&1.42&1.42\\
S1--C4&1.72&\\
$\angle$(C3-C1-C2)&113.31&113.3\\
$\angle$ (C2-C4-C5)&128.89&128.9\\
\toprule\\
\end{tabular}\\
 $^{a}$Data taken from from Ref. [23] 
\end{table}
The G$_0$W$_0$ approximation \cite{Ondia, Hedin} provided by the YAMBO code \cite{Marini} using TM pseudopotentials. The dielectric function is calculated using the plasmon-pole approximation, while the convergence in the self-energy and the dielectric matrix with respect to the number of bands are considered. The calculations were done with up to 200 empty bands corresponding to a maximal band energy of 0.44~a.u. (12 eV), as well as a cut-off for the dielectric matrix up to 2.5 a.u.
\section{RESULTS and DISCUSSIONS}
\subsection{Isolated polymer}
\begin{figure}
\caption {(Color online)  DFT-LDA (top) and G$_0$W$_0$ (down) band structure of polythiophne (PT). The Fermi energies are set to zero.}\label{fig3}
\hspace*{-0.6cm}
\centering
\includegraphics[scale=0.5]{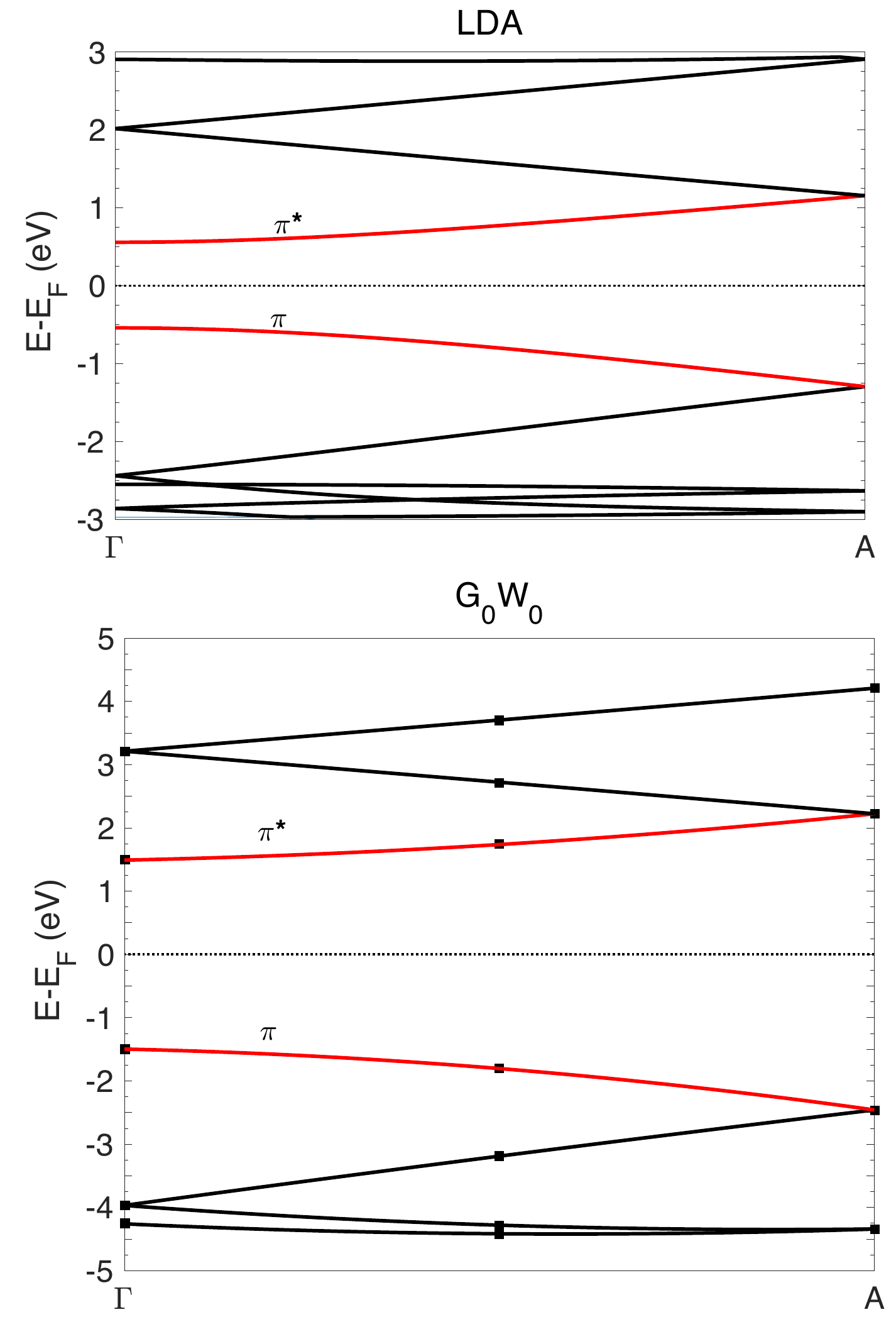} 
\end{figure}

Before considering the electronic properties of PT-absorbed on the graphene surface, we present the LDA band structure of isolated PT chain computed in the supercell shown in Fig \ref{fig1}(b). The polymer structure is constructed from the oligomer in the cell, when subjected to the PBC in both two in-plane directions, as shown in Fig. \ref{fig1}(a). The oligomer including four thiophene rings, as shown in Fig. \ref{fig2}, and is oriented parallel to the supercell base-diameter. The LDA optimized structure and geometry parameters are shown in Fig.s \ref{fig1}(b) and \ref{fig2}, respectively. The value of the corresponding parameters  are listed in Table \ref{tab1}. The obtained values are also compared with the previous theoretical LDA data, and the agreements support our technical process. By subjecting the supercell to the PBC, in fact we are neglecting the possible torsion angle between oligomers. By this assumption, the alignment of the $\pi$--orbitals will be maximized\cite{Scherf}, which is an appropriate model for the chains including more than 10 monomers \cite{Pesant}. 
\\
The LDA electronic band structure of the isolated PT are shown in Fig 3. Both the maximum of the valence and the minimum of the conduction bands are happen at the $\Gamma$ point. These two bands are stemming from the inter-ring $\pi$--bonding and $\pi^*$--antibonding states occur at -3.65 and -2.55 eV, respectively, and therefore, a direct gap of 1.10 eV is opened at the $\Gamma$ point. Depending on the pseudopotential and other calculation parameters given in Sec. \ref{sec: computational}, this gap is about 0.12-0.26 eV smaller than the previous LDA reported gaps \cite{Horst, Pesant}. The LDA valence and conduction bandwidth between point $\Gamma$ to the point A are found relatively small, as shown in Fig. \ref{fig3}, with value of 0.76 and  0.60 eV, respectively, and reflecting the wave functions are localized on individual chain.
The LDA band gaps obtained in this study and the previous efforts display relatively large deviation from experimental value (2.1 eV) \cite{Kobayashi}. One of the reasons is the spurious self-interaction error (SIE) related to the delocalization error of DFT exchange-correlation (xc) functionals stemming from dominating Coulomb term that pushes electrons apart \cite{Mori}. In some cases, hybrid functionals can partially correct for the SIE by including a fraction of exact exchange term adjusted to cancel the SIE \cite{Pesant}. However, using hybrid functionals is not a general solution and for a large number of atoms is computationally quite expensive. Here, we have performed DFT+U as an alternative method to cure the SIE by applying the U correction term only to the p orbitals. To determine parameters U and J, we follow a semi-empirical strategy and controlling the band gap. By increasing U and J to respectively 6 and 0.6 eV, the band gap is raised up and saturated at about 1.40 eV, which is not a significant improvement. Therefore, applying DFT+U, inspired by the Hubbard model, cannot provide a realistic model to correct the PT band gap, as has been proved in past for many conjugated polymers with delocalized interring $\pi$ electrons \cite{Baeriswyl}.\\
On the other hand, it has been shown that many body perturbations theory based on ab initio calculations of the quasiparticle (QP) band structure with the GW approximation overcomes the problematic effect of self-energy on polymer band gap \cite{Puschnig, Sun, Lanzillo, Samsonidze}. Therefore, we constructed GW calculations in a non-self-consistently manner from the LDA orbitals and eigenvalues, which is referred as the G$_0$W$_0$ method. The G$_0$W$_0$ approach is computationally so expensive for the present supercell introduced in Fig. \ref{fig1}(b). The sever limitation is the large number of atoms necessary to study the adsorption of PT on graphene surface. Therefore, we had to calculate electron self-energy by summing over computationally reasonable number of empty states. The initial calculation was performed with 100 number of unoccupied orbitals. The SIE corrected PT gap improved from LDA predicted 1.10 eV to G$_0$W$_0$ 2.96 eV at point $\Gamma$, as seen in Fig. \ref{fig3}. By increasing the number of empty states up to the 200 states, the energy of states are modified, however the initial G$_0$W$_0$ gap improves only by about 0.03 eV to 2.99 eV. Indeed, compared to LDA, the G$_0$W$_0$ predict larger bandwidth for the valence and conduction bands with value of 0.96 and  0.74 eV, respectively. 
Further increasing of the empty states in our G$_0$W$_0$ calculations imposed  computationally sever challenges, especially when the adsorption of PT on graphene is considered. The previous G$_0$W$_0$ gap computed within the generalized plasmon pole model and by summing self-energy over 1592 empty states within a smaller supercell including two thiophene rings is 3.10 eV, which is in satisfactory agreement with our outcomes driven from 200 empty states \cite{Samsonidze}. Here we point that the band gap of 3.59 eV is also predicted by a previous higher level calculation based on the self-consistent GW calculations \cite{Horst}. Moreover, the computed G$_0$W$_0$  quasiparticle gaps computed in this and the previous works are about 1 eV larger than the experimental value. The disagreement has been mostly cured  by including electron-hole interaction for the optical response, and inter-chain interactions in bulk polymer \cite{Horst, Samsonidze}.
\subsection{Uncovered graphene}\label{sec:	Uncovered graphene}
\begin{figure}
\caption{\label{fig4} {(Color online)  DFT-LDA band structure of isolated graphene calculated in the supercell described in the text. The Fermi energy is set to zero. The four touching bands forming two Dirac cones at $\Gamma$ are specified by red colors.}}
\hspace*{-0.6cm}
\centering
\includegraphics[scale=0.45]{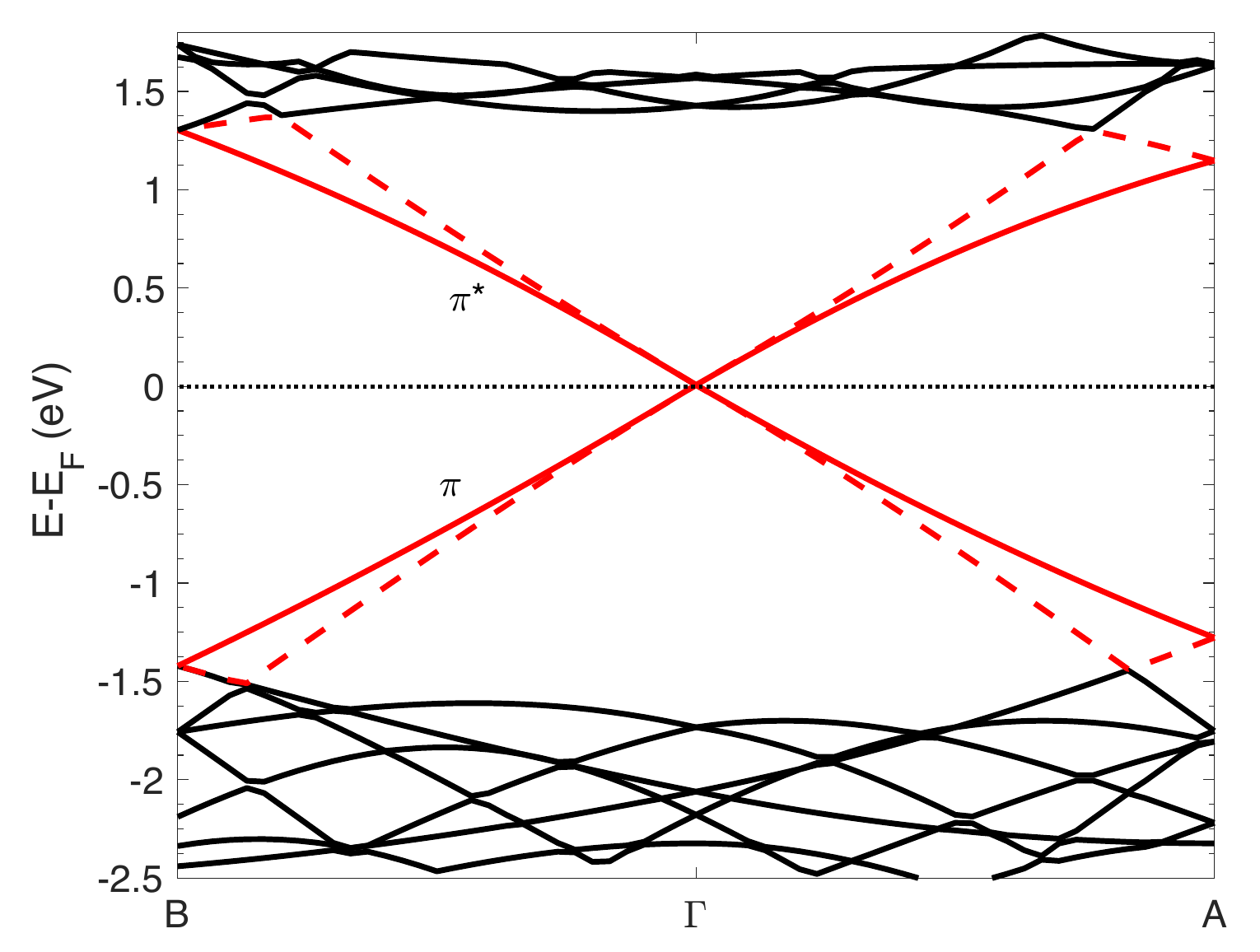} 
\end{figure}

At first, all carbon atoms in the graphene layer are relaxed to the ground state minimum potential energy point. The LDA predicts  C-C bond length of 1.429~\AA ~for graphene. Fig. \ref{fig4}  shows the LDA predicted band structure of pristine graphene deduced from the monoclinic supercell illustrated in Fig. \ref{fig1}(a) including 36 primitive cells and 72 carbons. Two of the six Dirac points at the corners of the hexagonal BZ are folded to  point $\Gamma$ inside the BZ shown in Fig. \ref{fig1}(c), so that there are four-fold degeneracy at this point with the energy of -2.13 eV, forming two pairs of touching cones. 
\subsection{PT adsorbed on graphene} \label{sec:	Uncovered graphene }
\begin{figure}
\caption{\label{fig5} {(Color online) Adsorption energy of PT on graphene surface obtained by using different functionals: GGA-PBE, LDA, and GGA-D2, which the latter includes a semi-empirical long-range dispersion correction, based on the Grimme$^\prime$s method, in terms of the adsorption distance. The zero line indicates the two independent components with no binding energy}}
\hspace*{-0.7cm}
\centering
\includegraphics[scale=0.45]{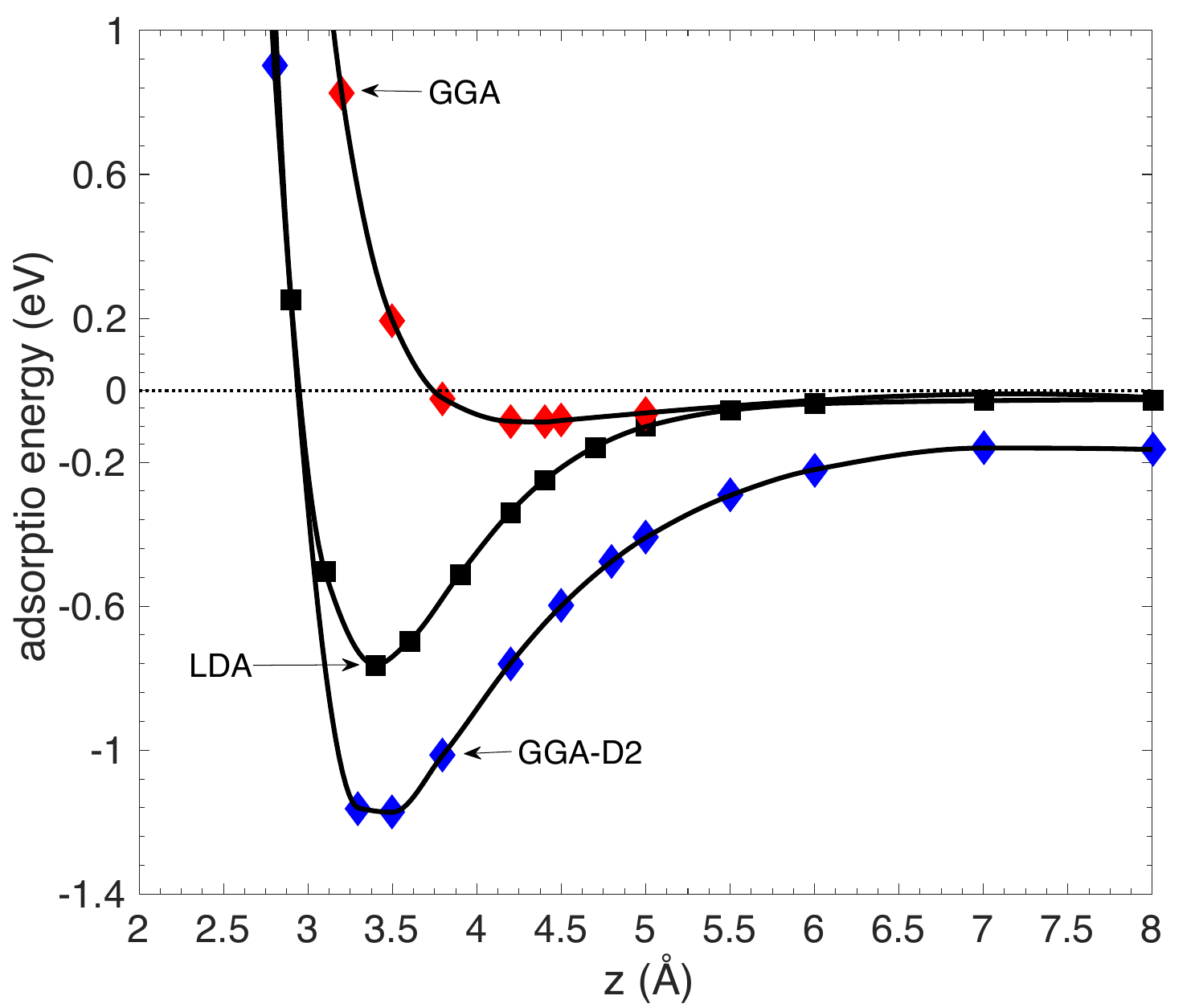} 
\end{figure}

The optimum distance between PT and graphene plane is determined by the value which minimized the total energy of the compound. PT adsorption energy on graphene surface as a function of adsorption distance, calculated by using different functionals are depicted in Fig. \ref{fig5}. In agreement with the previous report for poly(para-phenylene) (PPP) adsorbed on graphene \cite{Puschnig}, we also observe that adsorption energy and distance are sensitive to the choice of the applied functional. Moreover, each of LDA and GGA functionals exhibits different picture. According to LDA, the most stable geometry occurs in the adsorption distance of z$_0$ =3.4 \AA~ with binding energy of 0.77 eV. GGA-PBE functional predicts a shallow potential well at 4.2 \AA~ with binding energy of 0.17 eV. Since a typical substrate-adsorbate spacing in a physisorption is more than 3 \AA, these results suggest a physisorption of PT on graphene. The GGA-PBE small binding energy is a result of DFT deficient description of the long-range dispersion forces. The calculated binding energy based on LDA may not be reliable, since LDA does not include the long-ranged vdW interactions \cite{Nabok}. In fact, the over binding effect observed in the LDA functional compensates this missing term, and  LDA benefits from the cancellation of errors \cite{Nabok}. For GGA-PBE, one should contribute a small nonlocal vdW dispersion force, which almost is the binding force for the most of the organic materials \cite{ Dion}. Including DFT-D2 correction within Grimme empirical scheme \cite{Grimme}, as implemented in the ABINIT code, improves PBE binding energy to 1.16 eV at the adsorption distance of z$_0$= 3.5~\AA. Therefore, the contribution of vdW interactions to the binding energy amounts to 1 eV and indicates a strong physisorption. 
The orbital overlap and vdW interaction rearrange the charge density, which in turn should influence the valance and conduction energy levels of the polymer, and hence the band gap. Normally, the degree of the substrate-adsorbate vdW interaction depends on the adsorbed distance, the polarizability of the substrate and the number and type of the atoms involved. The reported vdW corrected GGA binding energy of PPP on the graphene surface is about 0.39 eV \cite{Puschnig}, which is so much smaller than what we obtained here for PT.\\

\begin{figure}
\caption{\label{fig6} {(Color online) Top panel: plane-averaged total charge density (red solid line) and charge density difference (black dashed line) of the combined system (G+PT) along the perpendicular direction to the graphene plane. Small charge modification are observed in the region between the graphene and polymer. Down panel: electrostatic potential energy  along the perpendicular direction to the graphene plane inside the supercell. The work function ($\phi$) of the combined system and uncovered graphene (G) are determined from the corresponding Fermi energies. . Work function at the bottom ( $\phi_b$) and the top sides ($\phi_t$) of the cell are illustrated. Black dashed line shows the electrostatic potential energy of the uncovered graphene. The dotted-black line indicates the Fermi energy of the uncovered graphene. The maximum value of the electrostatic potential energy of the combined system is set to zero.}}
\hspace*{- 1cm}
\centering
\includegraphics[scale=0.5]{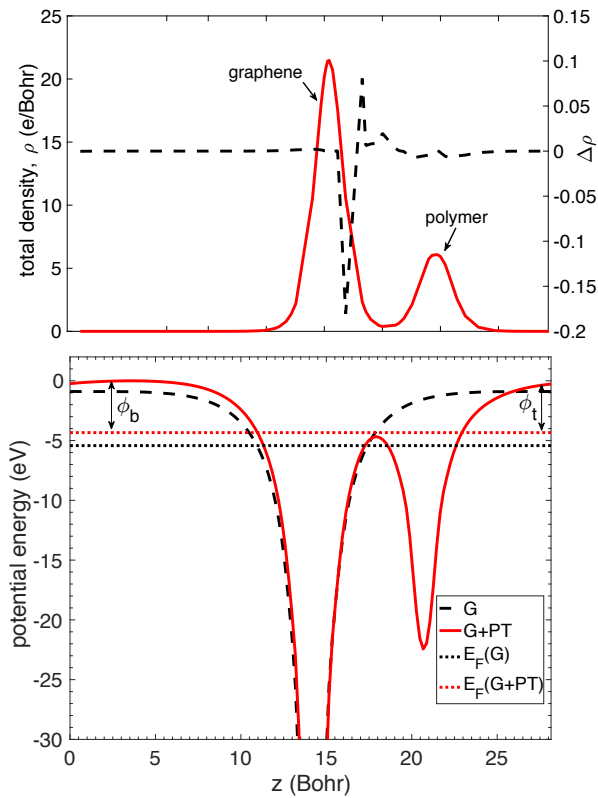} 
\end{figure}

\begin{figure}
\caption{\label{fig7} {(Color online) Top panel: LDA-predicted band structure of the combined system at an adsorption height of 3.4 ~\AA. The LDA valence and conduction bands of the isolated PT (blue-dashed lines) are shown to identify the corresponding states of the polymer (red-solid line) in the combined system. Zero energy is set to the Fermi energy of the combined system. Down panel: G$_0$W$_0$-predicted band structure of the combined system at an adsorption height of 3.4 ~\AA. The G$_0$W$_0$ valence and conduction bands of the isolated PT (blue-dashed line) are shown to identify the corresponding states of the polymer (red-solid line) in the combined system.  The G$_0$W$_0$ calculations were done in the three points shown by the square mark in each state. Zero energy is set to the energy of the graphene Dirac point (black-solid line at $\Gamma$ point) in the combined system.  
3.29 eV. The Fermi energy of the combined system is illustrated in dotted line of the lower panel. The Fermi energy of the uncovered graphene (at -2.13 eV) is not shown for clarity purpose and obtained 3.50~eV }}
\hspace*{-0.8cm}
\centering
\includegraphics[scale=0.5]{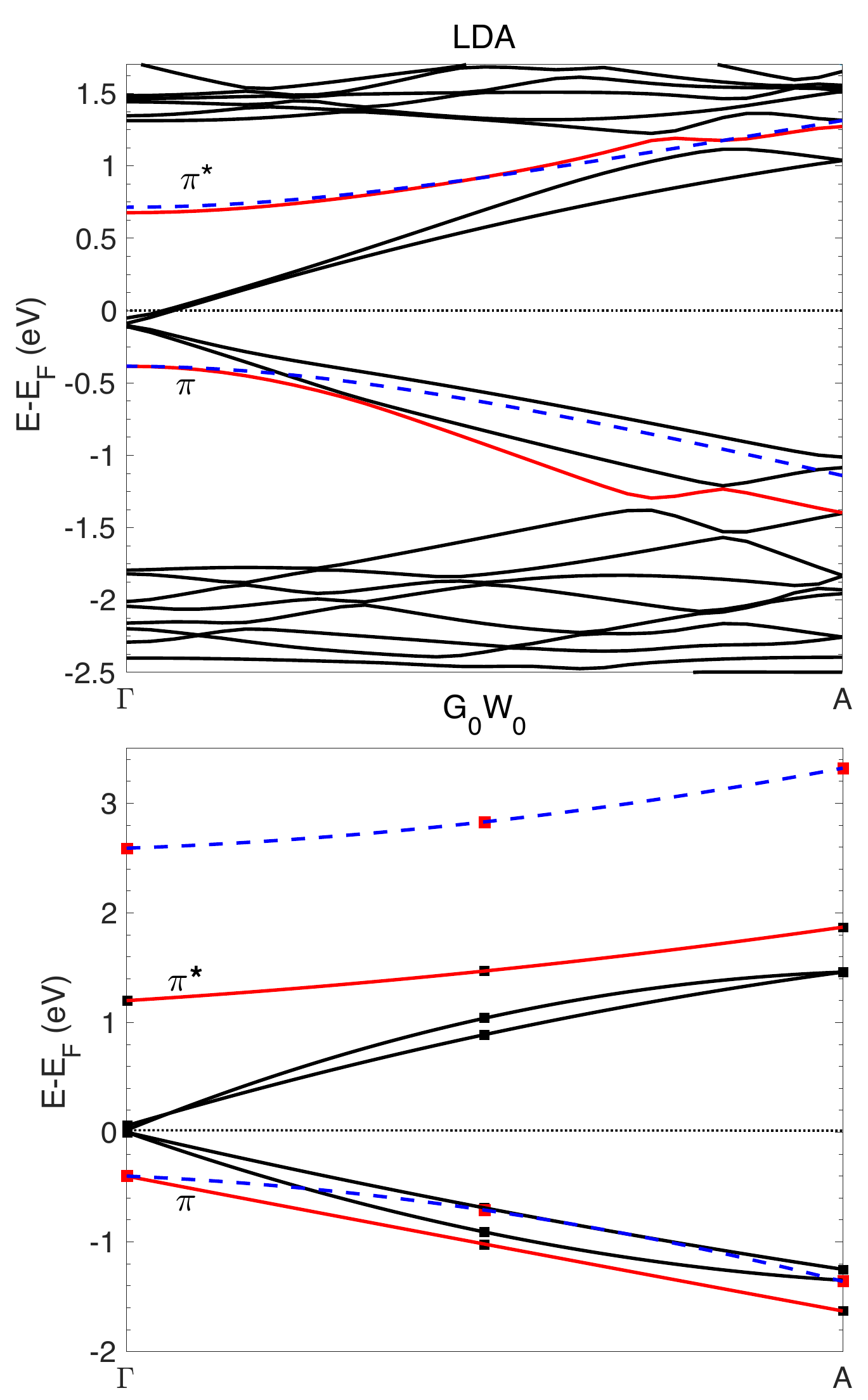} 
\end{figure}

To get more insight into the PT physisorption mechanism on graphene, we calculated the charge density rearrangement at the substrate-adsorbate interface by computing the in plane-averaged charge density of combined system relative to the charge densities of isolated components: $\Delta\rho=\rho$(G+PT)-($\rho$(PT)+$\rho$(G)). Here, $\rho$(G+PT) represents the charge density of the combined system, and $\rho$(PT), $\rho$(G) are the densities of the seperated polymer and uncovered graphene, respectively. The total plane-averaged and difference charge density, as well as the electrostatic potential are represented in Fig. \ref{fig6}. As shown in the top panel of Fig. \ref{fig6}, there is almost no change in the charge density of the polymer, suggesting no charge transfer between graphene and PT. The calculated charge density difference near the graphene surface obtained in this study is about one order of magnitude larger than for the PPP on graphene, reported in Ref. [13]. This result indicates the key role of the sulphur atom in this study, and explains the origin of the strong physisorbtion of PT on graphene. By considering $\Delta\rho$, we find electron charge density of graphene is attracted toward PT and makes a region of charge accumulation ($\Delta\rho >0$) and a region of charge depletion ($\Delta\rho >0$). Therefore the interaction between polymer and graphene gives rise to an interfacial electric dipole moment formation near the graphene surface in the region between two components. By PT absorption, the Fermi energy of the graphene exhibits a 1.09 eV downward shift, and the interfacial electric dipole moment makes a modification to graphene work function by a -0.19 eV shift down from its initial value of 4.53 eV (4.34~eV).\\
LDA band structure of PT at the 3.4~\AA ~above the graphene sheet is shown in Fig. \ref{fig7}. To identify the PT valence and conduction bands within the states of the combined system, the corresponding states of the isolated PT are appended to this figure. We see that according to LDA, PT band gap is slightly renormalized by a value of about 50 meV. The conduction band at point $\Gamma$ experiences only a small downward shift of 40 meV, which is the result of the weak charge density variation in PT after adsorption. As the result of the orbital hybridization, from the $\Gamma$ point toward point A, energy of the PT states are influenced by the graphene orbitals. \\

While many physical effects can influence the energy level of a molecule near a polarizable substrate, LDA (DFT) is not a successful tool to describe them.  An important effect in a physisorption is the Coulomb interaction between the surface and the electron or hole corresponding to the occupied or unoccupied states of the adsorbed molecule. The result of this interaction may cause the surface to polarize. The additional interaction between the polarized surface and the molecule influences the energy levels of the molecule and may strongly renormalize its gap.\\
 In classical picture, this subject can be described by a point charge \textit{q} located at the position $z_0$ above a polarizable surface with dielectric $\epsilon$ filling the half space $z<z_0$. The electrostatic interaction potential is given by \cite{White}\\
 \begin{equation}\label{eq.1}
V=\frac{qq^\prime}{4(z-z_0)},
\end{equation}
where the size of the image charge is $q^\prime =q(1-\epsilon)/(1+\epsilon)$. As a result, the energy of the unoccupied states should shift downward, since they experience an attractive potential according to Eq. \ref{eq.1}. For the occupied states the situation is inverse. Since the interaction with the positive image charge reduces the binding of the electron to the molecular core, these states are shifted upward. In contrast to LDA, long-range image static potential effects, for an electron near an interface can be well described by GW approximation. GW studies confirm that the electronic energy levels of a molecule outside a surface would also obey the image potential of Eq. \ref{eq.1}, and the response of the surface to an added electron or hole is reproduced within the G$_0$W$_0$ approach by means the screened Coulomb potential W$_0$. \cite{Puschnig, Fu, Neaton, Garcia, Rohlfing}. The interaction with the image charge shift the unoccupied levels downward, whereas the occupied levels upward and therefore, reduces the band gap.\\
The band structure of the combined PT-graphene system at the G$_0$W$_0$ level and adsorption distance of 3.4 \AA~ is shown in Fig. \ref{fig7}. For clarity, the valence and conduction bands of the combined system are specified in the red-solid lines and black squares. The corresponding isolated PT bands are also integrated to this figure and depicted by blue-dashed lines and red squares. In contrast to the LDA prediction, image charges induced in the graphene layer significantly perturb PT electronic states, so that the PT band gap reduces from 2.99 eV into 1.60 eV ($\Delta$E=1.39 eV). Four fold degeneracy of the graphene layer at the $\Gamma$ point is split by the vdW interaction, so that a gap of 60 meV is opened at this point.\\

\begin{figure}
\caption{\label{fig8} {(Color online) The electronic energy-level alignments of PT polymer obtained from DFT-LDA and G$_0$W$_0$ calculations near the combined graphene--PT interface in two different adsorption distances (4 and 3.4~\AA). The interface position is shown by red rectangular and the energy levels of polymer by blue line segments. The positions of each valence and conduction bands at points $"\Gamma"$ and $"A"$ are shown in eV. The vacuum level is set to 0 eV, and distances between line segments are scaled arbitrarily.}}
\hspace*{-0.7cm}
\centering
\includegraphics[scale=4]{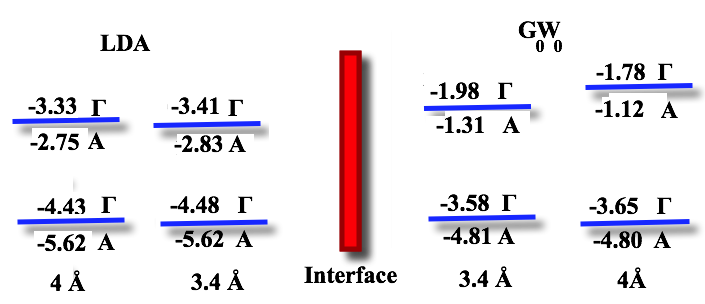} 
\end{figure}
We also study the dependence of the band alignment at the interface on the adsorption distance . Due to computational cost, our calculations are limited only to two cases with 3.4 \AA~ and 4 \AA~ distances from graphene layer. The results are shown in Fig. \ref{fig8}, which we plot the LDA and G$_0$W$_0$ energies of the valence and conduction bands of PT at $\Gamma$ and A points of BZ. Since the LDA respective valence and conduction positions of PT shift slightly in the same direction, the change in the gap is so small (30 meV at $\Gamma$ point). At the G$_0$W$_0$ level and from 4 into 3.4 \AA, the valence band moves up by 70 meV, whereas the conduction band moves down by 200 meV. Consequently, the band gap is reduced by 270 meV. Therefore, G$_0$W$_0$ not only can describe the band gap renormalization upon the dynamic polarization of the substrate, but also it can describe the effect of the substrate-adsorbate distance.
\section{Conclusion}

In summary, by local density approximation within DFT frame work and many-body perturbation theory at the G$_0$W$_0$ level, we investigate electronic properties and possible modifications of the electronic band structure of polythiophene by adsorbtion on graphene surface. The calculated charge density difference predicts the formation of an interface electric dipole near graphene surface, however we found that the charge transfer between them is negligible. Analysis of the electrostatic potential along the direction perpendicular to the graphene plane shows a -0.19 eV shift in the graphene work function from its initial value of 4.53 eV. According to LDA, the adsorption of PT on graphene does not alter the PT band gap, but orbital overlapping of PT and graphene modify the PT energy levels lineshapin far from the $\Gamma$ point. By reducing the substrate-adsorbate distance, the LDA valence and conduction bands move slightly in the same direction leading to a small change in the gap. However, by the G$_0$W$_0$ the valence band moves up and the conduction band moves down, results in a significant band reduction effect. According to our finding in this study, we predict a significant change of the polythiophene electronic behaviours in the presence of graphene.
\section{Acknowledgement}
F. Marsusi appreciates the computer assistance provided by Mrs Z.
Zeinali in the Department of Energy Engineering and Physics at
Amirkabir University of Technology. I. Fedorov gratefully acknowledge the Center for collective use “High Performance Parallel Computing” of the Kemerovo State University for providing the computational facilities.


\end{document}